\documentclass[final,1p,times]{elsarticle} 

\usepackage{amssymb}
\newcommand {\bc}{\begin{center}}
\newcommand {\ec}{\end{center}}
\def\lsim{\mathrel{\rlap{\lower4pt\hbox{$\sim$}}
    \raise1pt\hbox{$<$}}}               
\def\gsim{\mathrel{\rlap{\lower4pt\hbox{$\sim$}}
    \raise1pt\hbox{$>$}}}  
\newcommand {\bea}{\begin{eqnarray}}
\newcommand {\eea}{\end{eqnarray}}
\newcommand {\be}{\begin{equation}}
\newcommand {\ee}{\end{equation}}
\usepackage{graphicx}
\journal{Nuclear Physics A}

\begin{document}

\begin{frontmatter}



\title{Bulk viscosity, chemical equilibration
and flow at RHIC}


\author{Thomas Sch\"afer and Kevin Dusling}

\address{Department of Physics, North Carolina State
University, Raleigh, NC 27695}

\begin{abstract}
We study the effects of bulk viscosity on $p_T$ spectra and 
elliptic flow in heavy ion collisions at RHIC. We argue 
that direct effect of the bulk viscosity on the evolution
of the velocity field is small, but corrections to the 
freezeout distributions can be significant. These effects
are dominated by chemical non-equilibration in the hadronic
phase. We show that a non-zero bulk viscosity in the range 
$\zeta/s\lsim 0.05$ improves the description of spectra and 
flow at RHIC.

\end{abstract}

%\begin{keyword}


%\end{keyword}

\end{frontmatter}

% \linenumbers

\section{Introduction}
\label{sec_intro}
%%%%%%%%%%%%%%%%%%%%%%%%%%%%%%%%%%%%%%%%%%%%%%%%%%%%%%%%%%%%%%%%%%%%%%%%%

 The observation of nearly perfect hydrodynamic flow is one of the 
central discoveries of the heavy program at RHIC, and a significant 
amount of effort is being devoted to a precise determination of the 
shear viscosity to entropy density ratio $\eta/s$. In this contribution
we will try to estimate the bulk viscosity $\zeta$ of the excited 
matter created at RHIC. 

 Bulk viscosity enters the equations of fluid dynamics as an 
additional contribution to the stress tensor, $\delta T^{\mu\nu}=
-\Delta^{\mu\nu}\zeta\partial_k u^k$. Here, $\Delta^{\mu\nu}=g^{\mu\nu}
+ u^\mu u^\nu$ is a projector on the fluid rest frame and $u^\mu$ is 
the velocity of the fluid. Comparing with the stress tensor of an ideal 
fluid, $ T^{\mu\nu}=(\epsilon+P)u^\mu u^\nu + Pg^{\mu\nu}$ where 
$\epsilon$ is the energy density and $P$ is the pressure, we observe 
that bulk viscosity reduces the pressure of an expanding 
fluid relative to its equilibrium value. In a heavy ion collision
this implies that bulk viscosity reduces the amount of radial 
flow. 

 Bulk viscosity also effects the spectra of produced particles. 
Particle spectra are computed by matching the stress tensor 
across the freeze-out surface. This yields the standard Cooper-Frye
formula
\be 
 E_p\frac{dN}{d^3p} = \frac{1}{(2\pi)^3}\int_\sigma 
f(E_p)p^\mu d\sigma_\mu\, , 
\ee
where $dN/d^3p$ is the spectrum of produced particles, $E_p$ is 
the single particle energy, and $\sigma$ is the freeze-out surface. 
The distribution function $f(E_p)=f_0(E_p)+\delta f(E_p)$ contains 
an equilibrium part $f_0$ and a viscous correction $\delta f$. 
In the case of bulk viscosity $\delta f$ is proportional to the 
expansion rate $\partial_k u^k$, but the overall magnitude and
dependence on energy is sensitive to the underlying non-equilibrium
reactions.

%%%%%%%%%%%%%%%%%%%%%%%%%%%%%%%%%%%%%%%%%%%%%%%%%%%%%%%%%%%%%%%%%%%%%%%%%
\section{Theories and models of the single particle spectra}
\label{sec_delf}
%%%%%%%%%%%%%%%%%%%%%%%%%%%%%%%%%%%%%%%%%%%%%%%%%%%%%%%%%%%%%%%%%%%%%%%%%

 The bulk viscosity only constrains a moment of $\delta f$, and 
determining the full functional form of the non-equilibrium 
distribution function requires a microscopic model or theory. 
The simplest model is based on the Boltzmann equation in the 
relaxation time approximation. In this approximation the 
complicated collision term in the Boltzmann equation is parameterized
in terms of a single collision time $\tau(E_p)$. Energy and momentum 
conservation restrict the functional form of $\tau(E_p)$ 
\cite{Dusling:2011fd}. The bulk viscosity of an ultra-relativistic
gas in the relaxation time approximation is \cite{Weinberg:1971mx}
\be 
 \zeta = 15 \left(\frac{1}{3}-c_s^2\right)^2 \eta\, , 
\ee
where $c_s$ is the speed of sound. The non-equilibrium distribution
function is of the form 
\be 
\delta f \sim f^0_p \frac{\eta}{sT} \frac{p^2}{T^2} 
  \left( \frac{1}{3} - c_s^2\right) (\partial \cdot u)\, . 
\ee
We observe that the bulk viscosity scales as the second power
of the conformal breaking parameter $(c_s^2-1/3)$, whereas
$\delta f$ scales as the first power. This implies that for
a nearly conformal fluid corrections to the freeze-out
distribution are typically more important than corrections
to the velocity fields. This conclusion does not depend
on taking the relativistic limit. In general, the conformal breaking
parameter is 
\be 
{\cal F} = \int \frac{d^3p}{E_p(2\pi)^3} 
 \left(\frac{p^2}{3} -c_s^2 E_p^2\right) 
  f_0(p)\left(1\pm f_0(p)\right)\, 
\ee
where $\pm$ corresponds to bosons/fermions. The factor 
${\cal F}$ vanishes in both the relativistic limit $E_p\sim p$, 
and in the non-relativistic limit  $E_p\sim m+p^2/(2m)$. We find
that $\zeta\sim {\cal F}^2$ and $\delta f\sim {\cal F}$.

%%%%%%%%%%%%%%%%%%%%%%%%%%%%%%%%%%%%%%%%%%%%%%%%%%%%%%%%%%%%%%%%%%%%%%%%%
\begin{figure}[t]
\bc
\includegraphics[width=0.55\hsize]{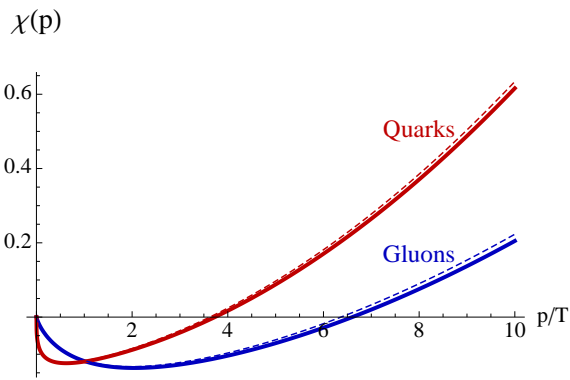}\ec
\caption{Non-equilibrium distribution $\chi$ of quarks and gluons 
in leading order perturbative QCD. The quantity $\chi$ is defined 
by $\delta f=-f_0(1\pm f_0)\chi(p)(\partial\cdot u)$. The dashed curves 
show an approximate solution that does not exactly conserve energy. }
\label{fig_delf}
\end{figure}
%%%%%%%%%%%%%%%%%%%%%%%%%%%%%%%%%%%%%%%%%%%%%%%%%%%%%%%%%%%%%%%%%%%%%%%%%

 The off-equilibrium distribution can be studied more rigorously 
in perturbative QCD \cite{Arnold:2006fz}. In QCD the process of 
emitting an extra soft gluon is efficient, and bulk viscosity 
is determined by the time scale for equilibrating the momenta
of produced gluons via elastic $2\leftrightarrow 2$ scattering. 
The off-equilibrium distribution of gluons with momenta $p\gg T$ 
is 
\be
 \delta f\sim f_p^0  \frac{p^2}{2\mu_A T} 
  \left( \frac{1}{3} - c_s^2\right) (\partial \cdot u)\, ,
\ee
where $\mu_A\sim g^2m_D^2\log(T/m_D)$ is the drag coefficient 
and $m_D$ is the Debye screening mass. In perturbative QCD 
$(c_s^2-1/3)=O(\alpha_s^2)$. In pure gauge theory $\zeta
=0.44\alpha_s^2 T^3/\log(\alpha_s^{-1})$ which implies $\zeta
\simeq 48 (c_s^2-1/3)^2\eta$. In full QCD we find interesting 
differences between the distribution functions of quarks and gluons, 
see Fig.~\ref{fig_delf}. Both off-equilibrium distribution functions
change sign, as is required by energy conservation, but the zero
crossing occurs for different momenta. 

 The difference between quarks and gluon distribution functions 
may manifest itself in viscous corrections to penetrating probes,
but more direct observables are related to the differences
between hadronic distribution functions. A problem that can be studied 
rigorously is the bulk viscosity of a pion gas \cite{Lu:2011df}. 
In this case bulk viscosity is determined by particle number 
changing processes, in particular the rate for the process
$\pi+\pi\leftrightarrow 4\pi$. In an expanding gas of massive
pions the equilibrium value of the total pion number decreases
with time, but pion number changing processes are slow and a 
pion excess is created. The distribution function can be parametrized
by an off-equilibrium chemical potential for the total number 
of pions
\be 
\label{del_f_pi}
\delta f = f_0 \left( \frac{\delta\mu}{T} 
  +  \frac{E_p\delta T}{T^2} \right) 
= -f_0 \left(\chi_0 - \chi_1 E_p \right)
  (\partial \cdot u)\, ,
\ee
where $\delta\mu$ is related to the bulk viscosity $\zeta$ and 
$\delta T$ is fixed by energy conservation. The bulk viscosity is 
controlled by the inelasticity $\Delta E=2m_\pi$, $\zeta \sim 
(f_\pi^8/m_\pi^5)\exp(-2m_\pi/T)$. We have extended equ.~(\ref{del_f_pi}) 
to a hadronic resonance gas, see \cite{Goity:1993ik} for earlier 
studies of chemical equilibration. We assume that 
the relative sizes of $\delta\mu$ for different species are determined 
by fast reactions like $\rho\leftrightarrow \pi\pi$ or $p+\bar{p}
\leftrightarrow n\pi$ ($n\simeq 5$ in the regime of interest).
This implies, for example, $\mu_\rho=2\mu_\pi$ and $\mu_p=2.5\mu_\pi$.
The overall scale of $\delta\mu$ is related to the bulk viscosity 
$\zeta$ and can be extracted from experiment. As before $\delta T$ 
is determined by energy conservation.

%%%%%%%%%%%%%%%%%%%%%%%%%%%%%%%%%%%%%%%%%%%%%%%%%%%%%%%%%%%%%%%%%%%%%%%%%
\section{Spectra and flow at RHIC}
\label{sec_flow}
%%%%%%%%%%%%%%%%%%%%%%%%%%%%%%%%%%%%%%%%%%%%%%%%%%%%%%%%%%%%%%%%%%%%%%%%%

%%%%%%%%%%%%%%%%%%%%%%%%%%%%%%%%%%%%%%%%%%%%%%%%%%%%%%%%%%%%%%%%%%%%%%%%%
\begin{figure} 
\bc\includegraphics[width=0.65\hsize]{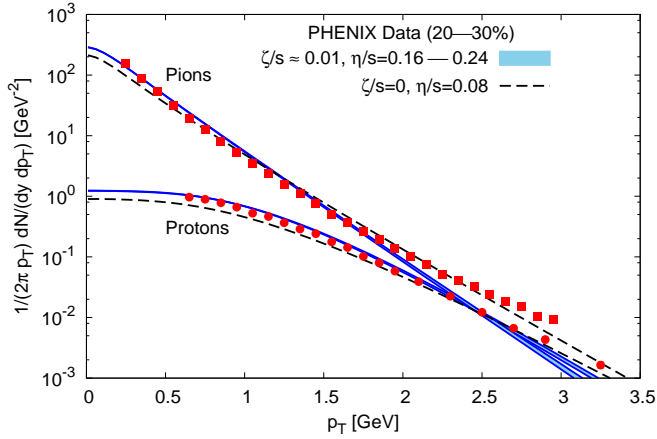}\ec
\caption{\label{fig_spectra}
Transverse momentum spectrum of pions and protons at RHIC
(data from \cite{Adler:2003cb}). The lines show hydrodynamic
calculations with and without bulk viscosity.}
\end{figure}
%%%%%%%%%%%%%%%%%%%%%%%%%%%%%%%%%%%%%%%%%%%%%%%%%%%%%%%%%%%%%%%%%%%%%%%%%

\begin{figure} 
\bc\includegraphics[width=0.65\hsize]{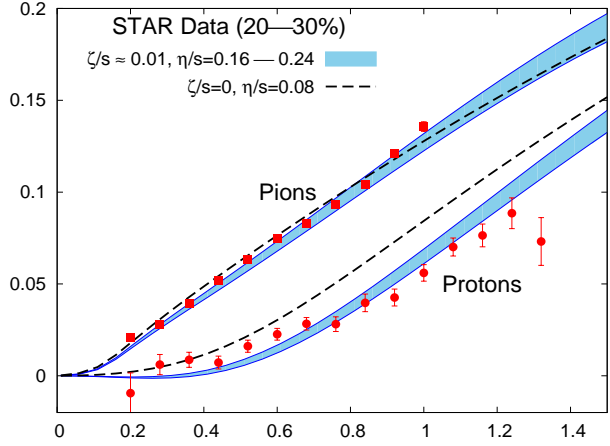}\ec
\caption{\label{fig_flow}
Differential elliptic flow of pions and protons at RHIC
(data from \cite{Adams:2004bi}) compared to hydrodynamic 
calculations with and without bulk viscosity. }
\end{figure}

 We have applied this model to $p_T$ spectra and flow at RHIC,
see Figs.~\ref{fig_spectra} and \ref{fig_flow}. The hydrodynamic
model incorporates a single freezeout temperature $T_{fr}\simeq
140$ MeV and no hadronic afterburner, but the spectra include
feed-down from hadronic resonances. The model for the temperature
dependence of the bulk viscosity is described in \cite{Dusling:2011fd}. 
The value of $\zeta/s$ quoted in the Figure refers to the freezeout
surface. We have solved second order 
hydrodynamic equations and we have verified that the gradient
expansion is convergent. Viscous corrections to the spectra 
become large for $p_T\gsim 2$ GeV, and results in this regime 
cannot be trusted. 
We observe that a non-zero
bulk viscosity improves the description of the spectra and flow. 
In particular, bulk viscosity raises the single particle spectra
at low $p_T$, and increases the splitting between the pion
and proton $v_2(p_T)$. These effects cannot be described in 
purely hydrodynamic models without bulk viscosity, but they 
have been explained in terms of hadronic non-equilibrium effects
in kinetic afterburners. Our results show that these effects
can be described efficiently in terms of bulk viscosity. We 
also note that many important hadronic reactions, such as 
$p\bar{p}$ annihilation into several pions, are difficult  
to include in kinetic models. Finally, we observe that the 
value of $\zeta$ extracted from the data is surprisingly 
small, $\zeta/s\simeq 0.01$. The corresponding pion chemical
potential is of the order $\mu_\pi\simeq (10-20)$ MeV, depending
on the local expansion rate.

%% T_{freezeout} =140 MeV
%% \tau_0=0.6 fm
%% e(x,y,\tau_0)=50 * n_{binary coll}
%% \sigma_{NN}=40
%% b=7.6 fm
%% Lat EoS
%% The bulk viscosity quoted in the legend is the approximate value at
%% Freezeout.  The bulk viscosity is given by eq. A9 in our paper where I
%% used sigma_0=0.025.

%%%%%%%%%%%%%%%%%%%%%%%%%%%%%%%%%%%%%%%%%%%%%%%%%%%%%%%%%%%%%%%%%%%%%%%%%
%%%%%%%%%%%%%%%%%%%%%%%%%%%%%%%%%%%%%%%%%%%%%%%%%%%%%%%%%%%%%%%%%%%%%%%%%

Acknowledgments: This work was supported in parts by the US 
Department of Energy grant DE-FG02-03ER41260. 



%% References


%\bibliographystyle{model1a-num-names}

\begin{thebibliography}{00}



\bibitem{Dusling:2011fd} 
  K.~Dusling and T.~Sch\"afer,
  %``Bulk viscosity, particle spectra and flow in heavy-ion collisions,''
  Phys.\ Rev.\ C {\bf 85}, 044909 (2012) 
  [arXiv:1109.5181 [hep-ph]].
  %%CITATION = ARXIV:1109.5181;%%

\bibitem{Weinberg:1971mx} 
  S.~Weinberg,
  %``Entropy generation and the survival of protogalaxies in an expanding universe,''
  Astrophys.\ J.\  {\bf 168}, 175 (1971).
  %%CITATION = ASJOA,168,175;%%

\bibitem{Arnold:2006fz} 
  P.~B.~Arnold, C.~Dogan and G.~D.~Moore,
  %``The Bulk Viscosity of High-Temperature QCD,''
  Phys.\ Rev.\ D {\bf 74}, 085021 (2006)
  [hep-ph/0608012].
  %%CITATION = HEP-PH/0608012;%%

\bibitem{Lu:2011df} 
  E.~Lu and G.~D.~Moore,
  %``The Bulk Viscosity of a Pion Gas,''
  Phys.\ Rev.\ C {\bf 83}, 044901 (2011)
  [arXiv:1102.0017 [hep-ph]].
  %%CITATION = ARXIV:1102.0017;%%

\bibitem{Goity:1993ik} 
  J.~L.~Goity,
  %``Chemical relaxation times in a hadron gas at finite temperature,''
  Phys.\ Lett.\ B {\bf 319}, 401 (1993);
  %[hep-ph/9310296].
  %%CITATION = HEP-PH/9310296;%%
%\bibitem{Pratt:1999ku} 
  S.~Pratt and K.~Haglin,
  %``Hadronic phase space density and chiral symmetry restoration in 
  %relativistic heavy ion collisions,''
  Phys.\ Rev.\ C {\bf 59}, 3304 (1999).
  %%CITATION = PHRVA,C59,3304;%%

%\bibitem{Bozek:2009dw} 
%  P.~Bozek,
%  %``Bulk and shear viscosities of matter created in relativistic 
%  %heavy-ion collisions,''
%  Phys.\ Rev.\ C {\bf 81}, 034909 (2010).
%  %[arXiv:0911.2397 [nucl-th]].
%  %%CITATION = ARXIV:0911.2397;%%

\bibitem{Adler:2003cb} 
  S.~S.~Adler {\it et al.}  [PHENIX collaboration],
  %``Identified charged particle spectra and yields in Au+Au collisions at S(NN)**1/2 = 200-GeV,''
  Phys.\ Rev.\ C {\bf 69}, 034909 (2004)
  [nucl-ex/0307022].
  %%CITATION = NUCL-EX/0307022;%%

\bibitem{Adams:2004bi} 
  J.~Adams {\it et al.}  [STAR collaboration],
  %``Azimuthal anisotropy in Au+Au collisions at s(NN)**(1/2) = 200-GeV,''
  Phys.\ Rev.\ C {\bf 72}, 014904 (2005)
  [nucl-ex/0409033].
  %%CITATION = NUCL-EX/0409033;%%

 \end{thebibliography}



\end{document}